\documentclass[english]{paper}
\usepackage[T1]{fontenc}
\usepackage[latin9]{inputenc}
\usepackage{float}
\usepackage{amsmath}
\usepackage{amssymb}
\usepackage{graphicx}
\usepackage{geometry}
\usepackage[numbers]{natbib}

\makeatletter
\let\paperclassexample\example

\let\example\relax
\AtBeginDocument{%
  \@ifundefined{example}{\let\example\paperclassexample}{}
}

\usepackage{booktabs}
\usepackage{amsmath,amsfonts,amsthm,bm}                
\usepackage{graphicx}
\usepackage{tikz}
\usetikzlibrary{matrix}
\usepackage{eso-pic}
\usepackage{amsmath}
\usetikzlibrary{tikzmark,fit}
\usepackage{tcolorbox}
\usepackage{nicematrix}

\makeatother

\usepackage{babel}
\begin{document}
\title{Dynamical Properties of the Multiple Elastic Pendulum.}
\author{Luis Rozas, luis.rozas@ulagos.cl{\huge\textbf{\date{}}}}
\institution{{\small Universidad de los Lagos, Departamento de Ciencias de la Ingeniería,
}\\
{\small Puerto Montt, Chile}}
\maketitle
\begin{abstract}
This article studies the dynamical properties of the multiple elastic
pendulum formed by an arbitrary number of point masses connected in
a chain by elastic springs. First, using the Lagrangian formulation,
the equations of motion governing the system's behavior, accounting
for both large-angle displacements and elastic deformations, are derived.
It is shown that well-established, low-dimensional systems, such as
the classical double elastic pendulum, can be systematically recovered
as particular cases directly from the general equations of motion
after applying the appropriates particular conditions. The behavior
of the system under small oscillations near its stable equilibrium
is also investigated. For this case, the linearized equations allow
the solution of the eigenvalue problem, demonstrating that in this
case uncoupled behavior between the angular displacements and the
spring elongations. 
\end{abstract}
\begin{keywords}
Lagrangian dynamics, multiple pendulum, linearization, eigenvalue
problem
\end{keywords}

\section{Introduction}

The study of the dynamical properties of pendulums has drawn the attention
of scientists and researchers for a long time. One of the first known
attempts to understand the behavior of such mechanical systems is
due to Galileo Galilei, who, after studying the motion of a swaying
lamp in the Pisa Cathedral, noticed that the time it took the lamp
to complete its swing was the same whether it traversed a larger or
smaller arc. This kind of simple pendulum with a stationary pivot
is entirely integrable: for small oscillations, using elementary trigonometric
functions, and for large oscillations, in terms of elliptic integrals
of the first kind and Jacobi elliptic functions \citep{Ochs2011,Belendez2012}. 

The double pendulum, which consists of a pendulum attached to the
free end of another pendulum, is a classic example of a simple mechanical
system that exhibits chaotic behavior. This pendulum has been widely
studied by several investigators, the first of whom was Daniel Bernoulli,
who published a paper on it in 1733 in the Proceedings of the Academy
of St. Petersburg. More recent studies of the double pendulum include
the work of A. Ohlhoff and P. H. Richter \citep{Ohlhoff2000}, which
focuses on the forces involved in its regular and chaotic behavior.
The double pendulum with distributed mass has been studied by Rafat
et al. \citep{Rafat2009}. Numerical analyses of the chaotic behavior
of the double pendulum---using a modified midpoint integrator, Poincaré
sections, and bifurcation diagrams---are presented in the work of
T. Stachowiak and T. Okada \citep{Stachowiak2006}.

More complex pendulum configurations have also been studied, such
as the triple pendulum in the works of Awrejcewicz et al. \citep{Awrejcewicz2007}
and Boran Yesilyurt \citep{Yesilyurt2020}. The double spring pendulum
with one rigid link has been studied by Wojciech Szumi\'{n}ski and
Andrzej J. Maciejewski \citep{Szuminski2024}, while the multi-pendulum
with $n$ point masses and rigid links is examined by M. Braun \citep{Braun2003}
and Boran Yesilyurt \citep{Yesilyurt2020}. 

The aim of this investigation is to derive the equations of motion
for a pendulum with $\ensuremath{n}$ point masses and elastic links,
and to analyze its dynamic properties under small oscillations.

\begin{figure}[H]
\centering\includegraphics[scale=0.55]{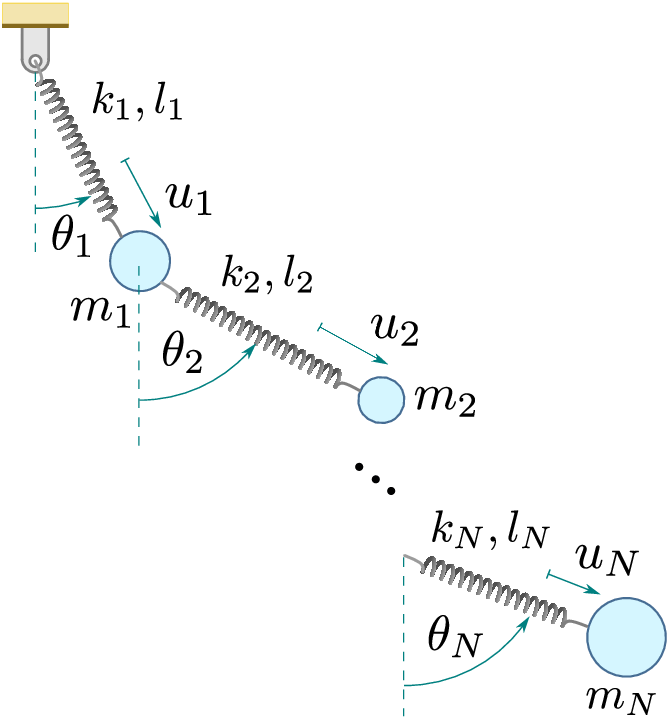}

\caption{Multiple elastic pendulum main parameters and generalized coordinates.\label{fig:MPendulum}}
\end{figure}

\section{Multiple elastic pendulum equations of motion }

As shown in Fig. \ref{fig:MPendulum}, the motion of the elastic pendulum
of $N$ point masses can be described by a set of $2N$ generalized
coordinates. These coordinates are the angles that each pendulum forms
with the vertical: $\theta_{1},\ldots,\theta_{N}$, and the deformation
of each spring: $u_{1},\ldots,u_{N}$, with respect to their natural
lengths: $l_{1},\ldots,l_{N}$. The equations of motion can be derived
by means of the Lagrange's equations \citep{Meirovitch1967} {\small
\begin{equation}
\frac{d}{dt}\left(\frac{\partial\mathcal{L}}{\partial\dot{q}_{i}}\right)-\frac{\partial\mathcal{L}}{\partial q_{i}}=Q_{i}\quad\quad i=1\ldots2N
\end{equation}
}where $\mathcal{L}=T-V$ is the Lagrangian, and $q_{i}$, $i=1\ldots2N$
is the set of generalized coordinates. For the evaluation of the kinetic
energy we need to find first the velocities of each mass. According
to Fig. \ref{fig:Velm_j}, the velocity of the j-th mass can be obtained
as the velocity of the first mass with respect to the fixed pivot,
plus the sum of the relative velocities of each mass relative to the
previous one, i.e.: $\boldsymbol{\upsilon}_{j}=\boldsymbol{\upsilon}_{1}+\sum_{i=2}^{j}\boldsymbol{\upsilon}_{i\mid i-1}$.
Therefore, the velocity of the j-th can be expressed in Cartesian
coordinates as:
\begin{figure}[H]
\centering\includegraphics[scale=0.75]{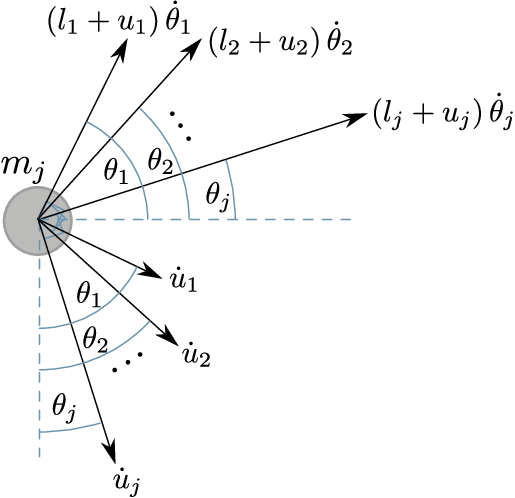}

\caption{Velocity of the j-th mass of the pendulum.\label{fig:Velm_j}}
\end{figure}
\begin{equation}
\boldsymbol{\upsilon}_{j}=\left\langle \sum_{i=1}^{j}\left(l_{i}+u_{i}\right)\dot{\theta}_{i}\cos\theta_{i}+\sum_{i=1}^{j}\dot{u}_{i}\sin\theta_{i},\sum_{i=1}^{j}\left(l_{i}+u_{i}\right)\dot{\theta}_{i}\sin\theta_{i}-\sum_{i=1}^{j}\dot{u}_{i}\cos\theta_{i}\right\rangle 
\end{equation}

The squared speed of the j-th mass is then:

\begin{equation}
\begin{array}{ccc}
\left\Vert \boldsymbol{\upsilon}_{j}\right\Vert ^{2} & = & {\displaystyle \sum_{i=1}^{j}}\left(l_{i}+u_{i}\right)^{2}\dot{\theta}_{i}^{2}\cos^{2}\theta_{i}+2{\displaystyle \sum_{1\leqslant r<s\leqslant j}}\left(l_{r}+u_{r}\right)\left(l_{s}+u_{s}\right)\dot{\theta_{r}}\dot{\theta_{s}}\cos\theta_{r}\cos\theta_{s}+\cdots\\
 &  & +{\displaystyle \sum_{i=1}^{j}}\left(l_{i}+u_{i}\right)^{2}\dot{\theta}_{i}^{2}\sin^{2}\theta_{i}+2{\displaystyle \sum_{1\leqslant r<s\leqslant j}}\left(l_{r}+u_{r}\right)\left(l_{s}+u_{s}\right)\dot{\theta_{r}}\dot{\theta_{s}}\sin\theta_{r}\sin\theta_{s}+\cdots\\
 &  & +{\displaystyle {\displaystyle \sum_{i=1}^{j}}}\dot{u}_{i}^{2}\sin^{2}\theta_{i}+2{\displaystyle \sum_{1\leqslant r<s\leqslant j}}\dot{u_{r}}\dot{u_{s}}\sin\theta_{r}\sin\theta_{s}+\cdots\\
 &  & +{\displaystyle \sum_{i=1}^{j}}\dot{u}_{i}^{2}\cos^{2}\theta_{i}+2{\displaystyle \sum_{1\leqslant r<s\leqslant j}}\dot{u_{r}}\dot{u_{s}}\cos\theta_{r}\cos\theta_{s}+\cdots\\
 &  & +2{\displaystyle \sum_{i=1}^{j}}\left(l_{i}+u_{i}\right)\dot{\theta}_{i}\dot{u_{i}}\sin\theta_{i}\cos\theta_{i}+{\displaystyle 2\sum_{\begin{array}{c}
1\leqslant r,s\leqslant j\\
r\neq s
\end{array}}}\left(l_{r}+u_{r}\right)\dot{\theta_{r}}\dot{u_{s}}\cos\theta_{r}\sin\theta_{s}+\cdots\\
 &  & -2{\displaystyle \sum_{i=1}^{j}}\left(l_{i}+u_{i}\right)\dot{\theta}_{i}\dot{u_{i}}\sin\theta_{i}\cos\theta_{i}-2{\displaystyle \sum_{\begin{array}{c}
1\leqslant r,s\leqslant j\\
r\neq s
\end{array}}}\left(l_{r}+u_{r}\right)\dot{\theta_{r}}\dot{u_{s}}\sin\theta_{r}\cos\theta_{s}
\end{array}
\end{equation}
using the appropriates trigonometric identities, the latter equation
can be simplified as:{\small
\begin{equation}
\begin{array}{ccc}
{\small \left\Vert \boldsymbol{\upsilon}_{j}\right\Vert ^{2}} & {\small =} & {\small {\displaystyle \sum_{i=1}^{j}}\left[\left(l_{i}+u_{i}\right)^{2}\dot{\theta}_{i}^{2}+\dot{u}_{i}^{2}\right]+2{\displaystyle \sum_{1\leqslant r<s\leqslant j}}\left[\left(l_{r}+u_{r}\right)\left(l_{s}+u_{s}\right)\dot{\theta_{r}}\dot{\theta_{s}}+\dot{u_{r}}\dot{u_{s}}\right]\cos\left(\theta_{r}-\theta_{s}\right)+\cdots}\\
 &  & +2{\displaystyle \sum_{\begin{array}{c}
1\leqslant r,s\leqslant j\\
r\neq s
\end{array}}}\left(l_{r}+u_{r}\right)\dot{\theta_{r}}\dot{u_{s}}\sin\left(\theta_{s}-\theta_{r}\right)
\end{array}
\end{equation}
}{\small\par}

Consequently, the total kinetic energy can be written as: $T=T_{a}+T_{b}+T_{c}$,
where the terms $T_{a}$, $T_{b}$ and $T_{c}$ are given by:
\begin{equation}
T_{a}=\frac{1}{2}\sum_{j=1}^{N}m_{j}{\displaystyle \sum_{i=1}^{j}}\left(\left(l_{i}+u_{i}\right)^{2}\dot{\theta}_{i}^{2}+\dot{u}_{i}^{2}\right)
\end{equation}

\begin{equation}
T_{b}=\sum_{j=1}^{N}m_{j}{\displaystyle \sum_{1\leqslant r<s\leqslant j}}\left[\left(l_{r}+u_{r}\right)\left(l_{s}+u_{s}\right)\dot{\theta_{r}}\dot{\theta_{s}}+\dot{u_{r}}\dot{u_{s}}\right]\cos\left(\theta_{r}-\theta_{s}\right)
\end{equation}

\begin{equation}
T_{c}=\sum_{j=1}^{N}m_{j}{\displaystyle \sum_{\begin{array}{c}
1\leqslant r,s\leqslant j\\
r\neq s
\end{array}}\left(l_{r}+u_{r}\right)\dot{\theta_{r}}\dot{u_{s}}\sin\left(\theta_{s}-\theta_{r}\right)}
\end{equation}

The term $T_{a}$ can be interpreted as the sum of the element of
the following lower triangular matrix:

\begin{equation}
T_{a}=\frac{1}{2}\mathbf{M} 
\begin{pNiceMatrix}
\left(l_{1}+u_{1}\right)^{2}\dot{\theta_{1}^{2}}+\dot{u}_{1}^{2}\\
\vdots & \ddots\\
\left(l_{1}+u_{1}\right)^{2}\dot{\theta_{1}^{2}}+\dot{u}_{1}^{2} & \cdots & \Block[fill=blue!20, rounded-corners]{3-1}{} \left(l_{k}+u_{k}\right)^{2}\dot{\theta_{k}^{2}}+\dot{u}_{k}^{2}\\
\vdots &  & \vdots & \ddots\\
\left(l_{1}+u_{1}\right)^{2}\dot{\theta_{1}^{2}}+\dot{u}_{1}^{2} & \cdots & \left(l_{k}+u_{k}\right)^{2}\dot{\theta_{k}^{2}}+\dot{u}_{k}^{2} & \cdots & \left(l_{N}+u_{N}\right)^{2}\dot{\theta_{N}^{2}}+\dot{u}_{N}^{2}
\end{pNiceMatrix}\label{eq:Ta_Matrix}
\end{equation}where $\mathbf{M}$ is the diagonal mass matrix of the pendulum. The
elements on the matrix of the equation \ref{eq:Ta_Matrix} that contains
the generalized coordinates and generalized velocities associated
with the index $k$ are shown in a shaded rectangle. With this in
mind, the partial derivatives of $T_{a}$ with respect to coordinates
$\theta_{k}$, $u_{k}$, and the velocities $\dot{\theta}_{k}$, $\dot{u}_{k}$
are: 
\begin{equation}
\left.\begin{array}{ccc}
{\displaystyle \frac{\partial T_{a}}{\partial\dot{\theta_{k}}}} & = & \left(l_{k}+u_{k}\right)^{2}\dot{\theta_{k}}{\displaystyle \sum_{r=k}^{N}m_{r}}\\
 & \vphantom{}\\
{\displaystyle \frac{\partial T_{a}}{\partial\theta_{k}}} & = & 0
\end{array}\right\} \label{eq:DTa_teta}
\end{equation}
\begin{equation}
\left.\begin{array}{ccc}
{\displaystyle \frac{\partial T_{a}}{\partial\dot{u_{k}}}} & = & \dot{u_{k}}{\displaystyle \sum_{r=k}^{N}m_{r}}\\
 & \vphantom{}\\
{\displaystyle \frac{\partial T_{a}}{\partial u_{k}}} & = & \left(l_{k}+u_{k}\right)\dot{\theta_{k}^{2}}{\displaystyle \sum_{r=k}^{N}m_{r}}
\end{array}\right\} \label{eq:DTa_u}
\end{equation}

The second term $T_{b}$ can be visualized as follows. All the elements
with index $j$ such as $j\geq k$ correspond to the sum of the elements
of the strictly upper matrix shown in equation \ref{eq:Tb_Matrix}:

\begin{equation}
T_{b}^{j\geq k}=m_{j} 
\begin{pNiceMatrix}

&0 & tb_{1,2} & \cdots & \Block[fill=blue!20, rounded-corners]{3-1}{} tb_{1,k} & \cdots & & tb_{1,j} \\
& &\ddots &\ddots &\vdots & & &\vdots  \\
&  & &0 & tb_{k-1,k} & \cdots &  & tb_{k-1,j}  \\
&  &  & &0 & \Block[fill=blue!20, rounded-corners]{1-3}{} tb_{k,k+1} & \cdots & tb_{k,j} \\
&  &  &  & & \ddots & \ddots & \vdots  \\
&  &  &  &  & &0 & tb_{j-1,j} \\
&  &  &  &  &  & &0 
\end{pNiceMatrix}\label{eq:Tb_Matrix}
\end{equation}where the notation: $tb_{r,s}=\left[\left(l_{r}+u_{r}\right)\left(l_{s}+u_{s}\right)\dot{\theta_{r}}\dot{\theta_{s}}+\dot{u_{r}}\dot{u_{s}}\right]\cos\left(\theta_{r}-\theta_{s}\right)$
is used for brevity. All elements in the matrix of the equation \ref{eq:Tb_Matrix}
that contains generalized coordinates and velocities associated with
the index $k$ are highlighted in the shaded rectangles. The sum of
these elements equals to:\footnote{Recall that: $tb_{r,s}=tb_{s,r}$ }:
\begin{equation}
\sum_{\begin{array}{c}
r=1\\
r\neq k
\end{array}}^{j}tb_{r,k}
\end{equation}

In consequence, the partial derivatives of $T_{b}$ with respect to
coordinates $\theta_{k}$, $u_{k}$, and the velocities $\dot{\theta}_{k}$,
$\dot{u}_{k}$, are shown in equations \ref{eq:DTb_Teta} and \ref{eq:DTb_u}:
\begin{equation}
\left.\begin{array}{ccc}
{\displaystyle \frac{\partial T_{b}}{\partial\dot{\theta_{k}}}} & = & {\displaystyle \sum_{j=1}^{N}m_{j}\sum_{\begin{array}{c}
r=1\\
r\neq k
\end{array}}^{j}}\left(l_{r}+u_{r}\right)\left(l_{k}+u_{k}\right)\dot{\theta_{r}}\cos\left(\theta_{r}-\theta_{k}\right)\\
 & \vphantom{}\\
{\displaystyle \frac{\partial T_{b}}{\partial\theta_{k}}} & = & {\displaystyle \sum_{j=1}^{N}m_{j}\sum_{\begin{array}{c}
r=1\\
r\neq k
\end{array}}^{j}}\left[\left(l_{r}+u_{r}\right)\left(l_{k}+u_{k}\right)\dot{\theta_{r}}\dot{\theta_{k}}+\dot{u_{r}}\dot{u_{k}}\right]\sin\left(\theta_{r}-\theta_{k}\right)
\end{array}\right\} \label{eq:DTb_Teta}
\end{equation}
\begin{equation}
\left.\begin{array}{ccc}
{\displaystyle \frac{\partial T_{b}}{\partial\dot{u_{k}}}} & = & {\displaystyle \sum_{j=k}^{N}m_{j}\sum_{\begin{array}{c}
r=1\\
r\neq k
\end{array}}^{j}}\dot{u_{r}}\cos\left(\theta_{r}-\theta_{k}\right)\\
 & \vphantom{}\\
{\displaystyle \frac{\partial T_{b}}{\partial u_{k}}} & = & {\displaystyle \sum_{j=k}^{N}m_{j}\sum_{\begin{array}{c}
r=1\\
r\neq k
\end{array}}^{j}}\left(l_{r}+u_{r}\right)\dot{\theta_{r}}\dot{\theta_{k}}\cos\left(\theta_{r}-\theta_{k}\right)
\end{array}\right\} \label{eq:DTb_u}
\end{equation}

The third term $T_{c}$ can also be understood as the sum of the elements
of a matrix. In this case the elements with index $j$ such as $j\geq k$
correspond to the sum of the elements of the hollow matrix shown in
equation \ref{eq:Tc_Matrix}:

\begin{equation}
T_{c}^{j\geq k}=m_{j} 
\begin{pNiceMatrix}
0 & \cdots & tc_{1,k-1} & \Block[fill=blue!20, rounded-corners]{3-1}{} tc_{1,k} & tc_{1,k+1} & \cdots & tc_{1,j} \\
\vdots & \ddots & \vdots & \vdots & \vdots & & \vdots\\
tc_{k-1,1} & \cdots & 0 & tc_{k-1,k} & tc_{k-1,k+1} & \cdots & tc_{k-1,j}\\
\Block[fill=blue!20, rounded-corners]{1-3}{} tc_{k,1} & \cdots & tc_{k,k-1} & 0 & \Block[fill=blue!20, rounded-corners]{1-3}{} tc_{k,k+1} & \cdots & tc_{k,j}\\
tc_{k+1,1} & \cdots & tc_{k+1,k-1} & \Block[fill=blue!20, rounded-corners]{3-1}{} tc_{k+1,k} & 0 & \cdots & tc_{k+1,j}\\
\vdots &  & \vdots & \vdots & \vdots & \ddots & \vdots\\
tc_{j,1} & \cdots & tc_{j,k-1} & tc_{j,k} & tc_{j,k+1} & \cdots & 0
\end{pNiceMatrix}\label{eq:Tc_Matrix}
\end{equation}here the notation $tc_{r,s}=\left(l_{r}+u_{r}\right)\dot{\theta_{r}}\dot{u_{s}}\sin\left(\theta_{s}-\theta_{r}\right)$
is used. Again, all elements in the matrix of equation \ref{eq:Tc_Matrix}
that contains generalized coordinates and velocities associated with
the index $k$ are highlighted in the shaded rectangles. The sum of
all these elements is: 
\begin{equation}
\sum_{\begin{array}{c}
r=1\\
r\neq k
\end{array}}^{j}tc_{k,r}+\sum_{\begin{array}{c}
r=1\\
r\neq k
\end{array}}^{j}tc_{r,k}\label{eq:SumTc_k}
\end{equation}

The partial derivatives of $T_{c}$ with respect to coordinates $\theta_{k}$,
$u_{k}$, and the velocities $\dot{\theta}_{k}$, $\dot{u}_{k}$ can
be obtained using the expression \ref{eq:SumTc_k}. The results are
shown in equations \ref{eq:DTc_Teta} and \ref{eq:DTc_u}: 

\begin{equation}
\left.\begin{array}{ccc}
{\displaystyle \frac{\partial T_{c}}{\partial\dot{\theta_{k}}}} & = & {\displaystyle \sum_{j=1}^{N}m_{j}\sum_{\begin{array}{c}
r=1\\
r\neq k
\end{array}}^{j}}\left(l_{k}+u_{k}\right)\dot{u_{r}}\sin\left(\theta_{r}-\theta_{k}\right)\\
 & \vphantom{}\\
{\displaystyle \frac{\partial T_{c}}{\partial\theta_{k}}} & = & {\displaystyle \sum_{j=1}^{N}m_{j}\sum_{\begin{array}{c}
r=1\\
r\neq k
\end{array}}^{j}}\left[\left(l_{r}+u_{r}\right)\dot{\theta_{r}}\dot{u_{k}}-\left(l_{k}+u_{k}\right)\dot{\theta_{k}}\dot{u_{r}}\right]\cos\left(\theta_{r}-\theta_{k}\right)
\end{array}\right\} \label{eq:DTc_Teta}
\end{equation}
\begin{equation}
\left.\begin{array}{ccc}
{\displaystyle \frac{\partial T_{c}}{\partial\dot{u_{k}}}} & = & {\displaystyle \sum_{j=k}^{N}m_{j}\sum_{\begin{array}{c}
r=1\\
r\neq k
\end{array}}^{j}}\left(l_{r}+u_{r}\right)\dot{\theta_{r}}\sin\left(\theta_{r}-\theta_{k}\right)\\
 & \vphantom{}\\
{\displaystyle \frac{\partial T_{c}}{\partial u_{k}}} & = & {\displaystyle \sum_{j=k}^{N}m_{j}\sum_{\begin{array}{c}
r=1\\
r\neq k
\end{array}}^{j}}\dot{\theta_{k}}\dot{\theta_{r}}\sin\left(\theta_{r}-\theta_{k}\right)
\end{array}\right\} \label{eq:DTc_u}
\end{equation}

On the other hand, the total potential energy of the pendulum can
be divided in the total gravitational potential energy: $U_{G}$,
and the total elastic energy stored by the springs: $U_{R}$. According
to Fig., the vertical displacement of the j-th mass of the pendulum
is: 
\begin{figure}[H]
\centering\includegraphics[scale=0.5]{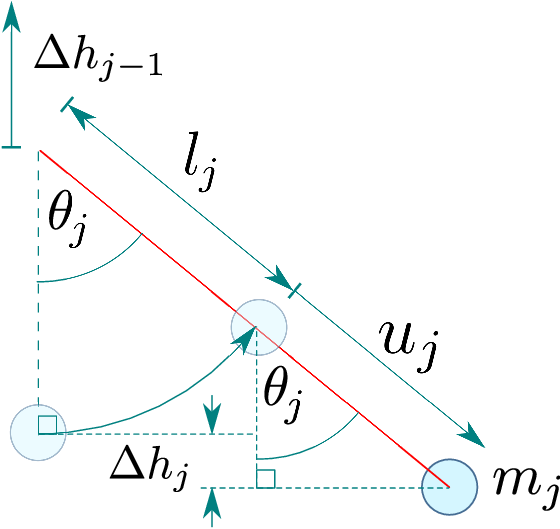}

\caption{Vertical displacement of the j-th mass.\label{fig:Pot_En}}
\end{figure}
\begin{equation}
\Delta h_{j}=\Delta h_{j-1}+\left(l_{j}-\left(l_{j}+u_{j}\right)\cos\theta_{j}\right)
\end{equation}

Thus:
\begin{equation}
U_{G}=g\sum_{j=1}^{N}m_{j}\sum_{i=1}^{j}l_{i}-\left(l_{i}+u_{i}\right)\cos\theta_{i}\label{eq:PotGrav}
\end{equation}

The total elastic energy stored by the springs is:
\begin{equation}
U_{R}=\dfrac{1}{2}\sum_{j=1}^{N}k_{j}u_{j}^{2}\label{eq:PotElas}
\end{equation}

Non conservative forces present in the system arise from different
sources. These can be represented as the forces exerted by equivalent
linear viscous dampers as follows:
\begin{equation}
Q_{\theta_{k}}=\left\{ \begin{array}{cc}
-c_{1}^{\theta}\dot{\theta_{1}}+c_{2}^{\theta}\left(\dot{\theta_{2}}-\dot{\theta_{1}}\right) & k=1\\
-c_{k}^{\theta}\left(\dot{\theta_{k}}-\dot{\theta}_{k-1}\right)+c_{k+1}^{\theta}\left(\dot{\theta_{k+1}}-\dot{\theta}_{k}\right) & k=1\ldots N-1\\
-c_{N}^{\theta}\left(\dot{\theta_{N}}-\dot{\theta}_{N-1}\right) & k=N
\end{array}\right.
\end{equation}

\begin{equation}
Q_{u_{k}}=-c_{k}^{u}\dot{u}_{k}
\end{equation}
where $c_{k}^{\theta}$ is the damping constant of the rotational
damper which connects the $k-1$ and $k$ spring, and $c_{k}^{u}$
is the damping constant of the extension damper of the $k$ spring. 

Using the results of equations: \ref{eq:DTa_teta}, \ref{eq:DTa_u},
\ref{eq:DTb_Teta}, \ref{eq:DTb_u}, \ref{eq:DTc_Teta}, \ref{eq:DTc_u}
and \ref{eq:PotElas}, and after evaluating the corresponding derivatives,
and simplify the results we obtain the equations of motion for the
multiple elastic pendulum. These equations can be arranged in two
groups. The first group contains the equations associated to the $\theta_{k}$($k=1\ldots N$)
coordinates, which are shown in the equation \ref{eq:Thetak}:{\small
\begin{equation}
\begin{array}{c}
\left(l_{k}+u_{k}\right)^{2}\sum_{j=k}^{N}m_{j}\ddot{\theta_{k}}+\sum_{j=k}^{N}m_{j}{\displaystyle \sum_{r=1,r\neq k}^{j}}\left(l_{r}+u_{r}\right)\left(l_{k}+u_{k}\right)\cos\left(\theta_{r}-\theta_{k}\right)\ddot{\theta_{r}}+\ldots\\
+\sum_{j=k}^{N}m_{j}{\displaystyle \sum_{r=1,r\neq k}^{j}}\left(l_{k}+u_{k}\right)\sin\left(\theta_{r}-\theta_{k}\right)\ddot{u_{r}}+2\left(l_{k}+u_{k}\right)\dot{u_{k}}\dot{\theta_{k}}\sum_{j=k}^{N}m_{j}+\ldots\\
+\sum_{j=k}^{N}m_{j}{\displaystyle \sum_{r=1,r\neq k}^{j}}\left[2\left(l_{k}+u_{k}\right)\dot{u_{r}}\dot{\theta_{r}}\cos\left(\theta_{r}-\theta_{k}\right)-\left(l_{r}+u_{r}\right)\left(l_{k}+u_{k}\right)\dot{\theta_{r}}^{2}\sin\left(\theta_{r}-\theta_{k}\right)\right]+\ldots\\
g\left(l_{k}+u_{k}\right)\sin\theta_{k}\sum_{j=k}^{N}m_{j}=Q_{\theta_{k}}
\end{array}\label{eq:Thetak}
\end{equation}
}{\small\par}

The second group, contains the equations associated to the $u_{k}$($k=1\ldots N$)
coordinates, and are shown in the equation \ref{eq:Uk}:

{\small
\begin{equation}
\begin{array}{c}
\sum_{j=k}^{N}m_{j}\ddot{u_{k}}+\sum_{j=k}^{N}m_{j}{\displaystyle \sum_{r=1,r\neq k}^{j}}\cos\left(\theta_{r}-\theta_{k}\right)\ddot{u_{r}}+\sum_{j=k}^{N}m_{j}{\displaystyle \sum_{r=1,r\neq k}^{j}}\left(l_{r}+u_{r}\right)\sin\left(\theta_{k}-\theta_{r}\right)\ddot{\theta_{r}}+\ldots\\
+\sum_{j=k}^{N}m_{j}{\displaystyle \sum_{r=1,r\neq k}^{j}}\left[2\dot{u_{r}}\dot{\theta_{r}}\sin\left(\theta_{k}-\theta_{r}\right)-\left(l_{r}+u_{r}\right)\dot{\theta_{r}}^{2}\cos\left(\theta_{k}-\theta_{r}\right)\right]-\left(l_{k}+u_{k}\right)\dot{\theta_{k}}^{2}\sum_{j=k}^{N}m_{j}+\ldots\\
+k_{k}u_{k}-g\cos\theta_{k}\sum_{j=k}^{N}m_{j}=Q_{u_{k}}
\end{array}\label{eq:Uk}
\end{equation}
}{\small\par}

The equations \ref{eq:Thetak} and \ref{eq:Uk} can be written in
compact matrix form as follows:{\small
\begin{equation}
\begin{array}{c}
\left[\begin{array}{cc}
\mathbf{M}_{\theta\theta} & \mathbf{M}_{\theta u}\\
\mathbf{M}_{u\theta} & \mathbf{M}_{uu}
\end{array}\right]\left\{ \begin{array}{c}
\ddot{\boldsymbol{\theta}}\\
\ddot{\mathbf{u}}
\end{array}\right\} +\left[\begin{array}{c}
\mathbf{N}_{\theta}\\
\mathbf{N}_{u}
\end{array}\right]\left\{ \dot{\boldsymbol{\theta}}\dot{\mathbf{u}}\right\} +\left[\begin{array}{c}
\mathbf{R}_{\theta}\\
\mathbf{R}_{u}
\end{array}\right]\left\{ \dot{\boldsymbol{\theta}}^{2}\right\} +\left[\begin{array}{cc}
\mathbf{C}_{\theta} & \mathbf{0}\\
\mathbf{0} & \mathbf{C}_{u}
\end{array}\right]\left\{ \begin{array}{c}
\dot{\boldsymbol{\theta}}\\
\dot{\mathbf{u}}
\end{array}\right\} +\ldots\\
\left[\begin{array}{cc}
\mathbf{0} & \mathbf{0}\\
\mathbf{0} & \mathbf{K}_{u}
\end{array}\right]\left\{ \begin{array}{c}
\boldsymbol{\theta}\\
\mathbf{u}
\end{array}\right\} +\left\{ \begin{array}{c}
\mathbf{p}_{\theta}\\
\mathbf{p}_{u}
\end{array}\right\} =\left\{ \begin{array}{c}
\mathbf{0}\\
\mathbf{0}
\end{array}\right\} 
\end{array}\label{eq:MatrixForm}
\end{equation}
}where $\boldsymbol{\theta}$ is the vector that contains the angular
coordinates, and $\mathbf{u}$ is the vector that contains the spring
deformations coordinates, calling $\mathbf{z}=\left\{ \begin{array}{cc}
\boldsymbol{\theta} & ,\mathbf{u}\end{array}\right\} ^{T}$, then:

\begin{equation}
\left\{ \dot{\boldsymbol{\theta}}\dot{\mathbf{u}}\right\} =\mathrm{diag}\left(P_{1}\dot{\mathbf{z}}\right)\cdot P_{2}\dot{\mathbf{z}}
\end{equation}
and
\begin{equation}
\left\{ \dot{\boldsymbol{\theta}}^{2}\right\} =\mathrm{diag}\left(P_{1}\dot{\mathbf{z}}\right)\cdot P_{1}\dot{\mathbf{z}}
\end{equation}
where $P_{1}=\left[\begin{array}{cc}
I_{N} & 0_{N}\end{array}\right]$, $P_{2}=\left[\begin{array}{cc}
0_{N} & I_{N}\end{array}\right]$ and $I_{N}$ and $0_{N}$ are the $N\times N$ identity and zero
matrix respectively. Therefore the vector $\left\{ \dot{\boldsymbol{\theta}}\dot{\mathbf{u}}\right\} $
correspond to the piece wise product of $\dot{\boldsymbol{\theta}}$
by $\dot{\mathbf{u}}$, and $\left\{ \dot{\boldsymbol{\theta}}^{2}\right\} $
is the vector that contains the squares of the angular velocities.
The expressions for the sub-matrices in \ref{eq:MatrixForm} are detailed
in Annex A. 

\subsection{Particular cases}

The equations \ref{eq:Thetak} and \ref{eq:Uk}, or equivalently the
matrix equation \ref{eq:MatrixForm}, can be used to obtain the equations
of motion for several interesting cases of pendulums. In the following
examples it is assumed that non conservatives forces are negligible
therefore: $Q_{\theta}=Q_{u}=0$. 

\subsubsection{Double pendulum}

For $N=2$ and in the rigid case $k_{1,2}\rightarrow\infty$, the
system reduces to the standard double pendulum. Under the latter condition,
$u_{1}$ and $u_{2}$ become constrained coordinates, and their values
as well as their time derivatives: $u_{1},u_{2},\dot{u_{1}},\dot{u_{2}},\ddot{u_{1}},\ddot{u_{2}}$
vanish identically. Substituting these constraints into equations
\ref{eq:Thetak} and \ref{eq:Uk} yields, after simplification, the
following equations of motion:{\small
\begin{equation}
\begin{array}{ccc}
\left(m_{1}+m_{2}\right)l_{1}\ddot{\theta_{1}}+m_{2}l_{2}\ddot{\theta_{2}}\cos\left(\theta_{2}-\theta_{1}\right)-m_{2}l_{2}\dot{\theta_{2}^{2}}\sin\left(\theta_{2}-\theta_{1}\right)+g\left(m_{1}+m_{2}\right)\sin\theta_{1} & = & 0\\
l_{2}\ddot{\theta_{2}}+l_{1}\cos\left(\theta_{2}-\theta_{1}\right)\ddot{\theta_{1}}+l_{1}\dot{\theta_{1}^{2}}\sin\left(\theta_{2}-\theta_{1}\right)+g\sin\theta_{2} & = & 0
\end{array}\label{eq:DoublePend}
\end{equation}
}{\small\par}

Equations \ref{eq:DoublePend} represent the equations of motion for
the double pendulum. Its behavior has been extensively analyzed in
the literature, including studies by:\citep{Stachowiak2006,Rafat2009,Ohlhoff2000}.

\subsubsection{Triple pendulum}

This case occurs when $N=3$ and $k_{1,2,3}\rightarrow\infty$. Similar
to the previous case here $u_{1}$, $u_{2}$ and $u_{3}$ and all
its derivatives are zero. The use of equations \ref{eq:Thetak} and
\ref{eq:Uk} give us in this case the following results
\begin{equation}
\begin{array}{lcc}
\underset{1\rightarrow3}{m}l_{1}\ddot{\theta_{1}}+\underset{2\rightarrow3}{m}l_{2}\cos\left(\theta_{2}-\theta_{1}\right)\ddot{\theta_{2}}+m_{3}l_{3}\cos\left(\theta_{3}-\theta_{1}\right)\ddot{\theta_{3}}+\ldots\\
\underset{2\rightarrow3}{m}l_{2}\sin\left(\theta_{1}-\theta_{2}\right)\dot{\theta_{2}^{2}}+m_{3}l_{3}\dot{\theta_{3}^{2}}\sin\left(\theta_{1}-\theta_{3}\right)+\underset{1\rightarrow3}{m}g\sin\theta_{1} & = & 0\\
\underset{2\rightarrow3}{m}l_{2}\ddot{\theta_{2}}+\underset{2\rightarrow3}{m}l_{1}\cos\left(\theta_{2}-\theta_{1}\right)\ddot{\theta_{1}}+m_{3}l_{3}\cos\left(\theta_{3}-\theta_{2}\right)\ddot{\theta_{3}}+\ldots\\
\underset{2\rightarrow3}{m}l_{1}\sin\left(\theta_{2}-\theta_{1}\right)\dot{\theta_{1}^{2}}+m_{3}l_{3}\sin\left(\theta_{2}-\theta_{3}\right)\dot{\theta_{3}^{2}}+g\underset{2\rightarrow3}{m}\sin\theta_{2} & = & 0\\
m_{3}l_{3}\ddot{\theta_{3}}+m_{3}l_{1}\cos\left(\theta_{1}-\theta_{3}\right)\ddot{\theta_{1}}+m_{3}l_{2}\cos\left(\theta_{2}-\theta_{3}\right)\ddot{\theta_{2}}+\ldots\\
m_{3}l_{1}\sin\left(\theta_{3}-\theta_{1}\right)\dot{\theta_{1}^{2}}+m_{3}l_{2}\sin\left(\theta_{3}-\theta_{2}\right)\dot{\theta_{2}^{2}}+gm_{3}\sin\theta_{3} & = & 0
\end{array}\label{eq:Triple_Pend}
\end{equation}
here: $\underset{i\rightarrow j}{m}=\sum_{r=i}^{j}m_{r}$. The equations
\ref{eq:Triple_Pend} govern the behavior of the triple pendulum.
Studies of this mechanical system include the work of Awrejcewicz
et al. \citep{Awrejcewicz2007} and Boran Yesilyurt \citep{Yesilyurt2020}

\subsubsection{Double spring pendulum, one rigid link}

In this case $N=2$ and $k_{1}\rightarrow\infty$, thus $u_{1}=\dot{u_{1}}=\ddot{u}_{1}=0$.
Using a similar procedure as in the previous cases, the following
are the equations of motion:
\begin{equation}
\begin{array}{lcc}
\left(m_{1}+m_{2}\right)l_{1}\ddot{\theta_{1}}+m_{2}\left(l_{2}+u_{2}\right)\cos\left(\theta_{2}-\theta_{1}\right)\ddot{\theta_{2}}+m_{2}\sin\left(\theta_{2}-\theta_{1}\right)\ddot{u_{2}}+\ldots\\
2m_{2}\dot{u_{2}}\dot{\theta_{2}}\cos\left(\theta_{2}-\theta_{1}\right)+m_{2}\left(l_{2}+u_{2}\right)\sin\left(\theta_{1}-\theta_{2}\right)\dot{\theta_{2}^{2}}+g\left(m_{1}+m_{2}\right)\sin\theta_{1} & = & 0\\
m_{2}\left(l_{2}+u_{2}\right)\ddot{\theta_{2}}+m_{2}l_{1}\cos\left(\theta_{2}-\theta_{1}\right)\ddot{\theta_{1}}+2m_{2}\dot{u_{2}}\dot{\theta_{2}}+\ldots\\
m_{2}l_{1}\dot{\theta_{1}^{2}}\sin\left(\theta_{2}-\theta_{1}\right)+gm_{2}\sin\theta_{2} & = & 0\\
m_{2}\ddot{u_{2}}+m_{2}l_{1}\sin\left(\theta_{2}-\theta_{1}\right)\ddot{\theta_{1}}-m_{2}l_{1}\cos\left(\theta_{2}-\theta_{1}\right)\dot{\theta_{1}^{2}}-m_{2}\left(l_{2}+u_{2}\right)\dot{\theta_{2}^{2}}+\ldots\\
k_{2}u_{2}-gm_{2}\cos\theta_{2} & = & 0
\end{array}
\end{equation}

These equations can also be found in the work of Wojciech Szumi\'{n}ski
and Andrzej J. Maciejewski \citep{Szuminski2024}.

\subsubsection{Multiple rigid pendulum}

For the multiple rigid pendulum $u_{i}=\dot{u_{i}}=\ddot{u_{i}}=0$,
$i=1\ldots N$. The resultant equations are: 
\begin{equation}
\begin{array}{c}
l_{k}^{2}\sum_{j=k}^{N}m_{j}\ddot{\theta_{k}}+\sum_{j=k}^{N}m_{j}{\displaystyle \sum_{r=1,r\neq k}^{j}}l_{r}l_{k}\cos\left(\theta_{r}-\theta_{k}\right)\ddot{\theta_{r}}+\ldots\\
+\sum_{j=k}^{N}m_{j}{\displaystyle \sum_{r=1,r\neq k}^{j}}l_{r}l_{k}\dot{\theta_{r}}^{2}\sin\left(\theta_{k}-\theta_{r}\right)+gl_{k}\sin\theta_{k}\sum_{j=k}^{N}m_{j}=0
\end{array}
\end{equation}

These equations are identical to those found in the works of M.Braun
\citep{Braun2003}, and Boran Yesilyurt \citep{Yesilyurt2020}.

\section{Dynamic behavior under small deformations}

Assuming all angles and spring elongations are small, the second-order
terms containing the squares of velocities and the products of coordinates
or velocities vanish. Additionally, we can apply standard trigonometric
approximations to obtain equivalent linear equations. These considerations
allow us to simplify the equations \ref{eq:Thetak} and \ref{eq:Uk}
as follows:
\begin{equation}
\begin{array}{ccc}
l_{k}^{2}\sum_{j=k}^{N}m_{j}\ddot{\theta_{k}}+\sum_{j=k}^{N}m_{j}{\displaystyle \sum_{r=1,r\neq k}^{j}}l_{r}l_{k}\ddot{\theta_{r}}+gl_{k}\theta_{k}\sum_{j=k}^{N}m_{j} & = & 0\\
\sum_{j=k}^{N}m_{j}\ddot{u_{k}}+\sum_{j=k}^{N}m_{j}{\displaystyle \sum_{r=1,r\neq k}^{j}}\ddot{u_{r}}+k_{k}u_{k}-g\sum_{j=k}^{N}m_{j} & = & 0
\end{array}\label{eq:Linear_eq}
\end{equation}

Its easy to see that in this situation the angles and the elongations
of the springs remains uncoupled. The equations \ref{eq:Linear_eq}
can be written in matrix form:
\begin{equation}
\begin{array}{ccl}
\mathbf{\mathbb{M}}_{\theta}\ddot{\boldsymbol{\theta}}+\mathbf{\mathbb{K}}_{\theta}\boldsymbol{\theta} & = & 0\\
\mathbf{\mathbb{M}}_{u}\ddot{\mathbf{u}}+\mathbf{\mathbb{K}}_{u}\mathbf{u} & = & g\overrightarrow{\mathrm{diag}\left(\mathbf{\mathbb{M}}_{u}\right)}
\end{array}\label{eq:Linear_eq_Matrix}
\end{equation}
and $\overrightarrow{\mathrm{diag}\left(\mathbf{\mathbb{M}}_{u}\right)}$
is the column vector formed by the diagonal of the matrix $\mathbf{\mathbb{M}}_{u}$.
The matrices in the equation \ref{eq:Linear_eq_Matrix} are defined
in Annex B. Using the change of variables: $\mathbf{u}=\overline{\mathbf{u}}+\mathbf{u}_{0}$,
where $\overline{\mathbf{u}}$ is the vector of springs elongations
relative to the static equilibrium position and $\mathbf{u}_{0}=\mathbf{\mathbb{K}}_{u}^{-1}\cdot g\overrightarrow{\mathrm{diag}\left(\mathbf{\mathbb{M}}_{u}\right)}$
is the vector of their static elongations, the equations \ref{eq:Linear_eq_Matrix}
can be written as:
\begin{equation}
\begin{array}{ccl}
\mathbf{\mathbb{M}}_{\theta}\ddot{\boldsymbol{\theta}}+\mathbf{\mathbb{K}}_{\theta}\boldsymbol{\theta} & = & 0\\
\mathbf{\mathbb{M}}_{u}\ddot{\mathbf{\overline{u}}}+\mathbf{\mathbb{K}}_{u}\overline{\mathbf{u}} & = & 0
\end{array}\label{eq:Linear_eq_Matrix_1}
\end{equation}

The equations \ref{eq:Linear_eq_Matrix_1} are the linearized equations
of the multiple elastic pendulum about the static equilibrium position.
Due to the fact that coordinates $\boldsymbol{\theta}$ and $\overline{\mathbf{u}}$
are uncoupled, the two sets of equations in \ref{eq:Linear_eq_Matrix_1}
can be analyzed independently. The first set of equations in \ref{eq:Linear_eq_Matrix_1}
dictates the behavior under small deformations of the multiple pendulum
with rigid links. This pendulum is studied in detail in the work of
M.Braun \citep{Braun2003}, particularly the eigenvalue problem leads
to the following important results regarding the natural frequencies
of oscillation, $\omega_{\theta}$, of the pendulum
\begin{equation}
\begin{array}{ccl}
\sum_{j=1}^{N}\frac{1}{\omega_{\theta}^{2}} & = & \frac{L}{g}\\
\prod_{j=1}^{N}\frac{1}{\omega_{\theta}^{2}} & = & \prod_{j=1}^{N}\frac{l_{j}}{g}\cdot\prod_{j=1}^{N-1}\frac{m_{j}}{_{j}^{N}m}
\end{array}
\end{equation}
where $L=\sum_{j=1}^{N}l_{j}$. Following a similar procedure to that
developed by M.Braun, the natural frequencies of oscillations associated
to the elongations of the springs can be found assuming that the solution
of the second set of equations in \ref{eq:Linear_eq_Matrix_1} is:
\begin{equation}
\mathbf{\overline{u}}=\phi_{j}q_{0_{j}}e^{i\omega_{uj}t}\label{eq:Modal_u}
\end{equation}

In the equation \ref{eq:Modal_u}, $\phi_{j}$ are the mode shape
and $\omega_{uj}$ are the natural frequencies of oscillation, $j=1\ldots N$.
Replacing \ref{eq:Modal_u} in the second set of equations in \ref{eq:Linear_eq_Matrix_1},
and assuming the homogeneous case we obtain:
\begin{equation}
\left(\mathbf{\mathbb{K}}_{u}^{-1}\mathbf{\mathbb{M}}_{u}\right)\phi_{j}=\frac{1}{\omega_{uj}^{2}}\phi_{j}
\end{equation}

This is the classic standard eigenvalue problem for the matrix $\mathbf{A}=\mathbf{\mathbb{K}}_{u}^{-1}\mathbf{\mathbb{M}}_{u}$,
with eigenvectors $\phi_{j}$ and eigenvalues $1/\omega_{uj}^{2}$.
The sum of the eigenvalues is equal to the trace of matrix $\mathbf{A}$,
therefore:
\begin{equation}
\sum_{j=1}^{N}\frac{1}{\omega_{uj}^{2}}=\sum_{j=1}^{N}\frac{_{j}^{N}m}{k_{j}}\label{eq:Sum_wu}
\end{equation}

In the case that all springs has the same stiffness, ie. :$k_{j}=k$,
$j=1\ldots N$, then the equation \ref{eq:Sum_wu} becomes:
\begin{equation}
\sum_{j=1}^{N}\frac{1}{\omega_{uj}^{2}}=\sum_{j=1}^{N}\frac{j}{\omega_{j}^{2}}
\end{equation}
where $\omega_{j}^{2}=k/m_{j}$. If also all masses are equal ($m_{j}=m$,
$j=1\ldots N$) then:
\begin{equation}
\sum_{j=1}^{N}\frac{1}{\omega_{uj}^{2}}=\frac{N}{2\omega^{2}}\left(N+1\right)
\end{equation}
where $\omega=k/m$. On the other hand, the determinant of the matrix
$\mathbf{A}$ give us the product of the eigenvalues. This determinant
can be obtained as the product of the diagonal elements of the Gaussian
upper triangular matrix obtained after applying forward Gaussian elimination.
The latter procedure leads us to the following result: 
\begin{equation}
\prod_{j=1}^{N}\frac{1}{\omega_{uj}^{2}}=\frac{_{1}^{N}m}{k_{1}}\prod_{j=1}^{N}\frac{_{j}^{N}m}{k_{j}}\left(1-\frac{_{j}^{N}m}{_{j-1}^{N}m}\right)
\end{equation}
which can be simplified as:
\begin{equation}
\prod_{j=1}^{N}\frac{1}{\omega_{uj}^{2}}=\prod_{j=1}^{N}\frac{m_{j}}{k_{j}}
\end{equation}

In the special case where: $m_{j}=m$, and $k_{j}=k$ ($j=1\ldots N$)
\begin{equation}
\prod_{j=1}^{N}\frac{1}{\omega_{uj}^{2}}=\left(\frac{1}{\omega}\right)^{2N}
\end{equation}

\section{Conclusions}

This article presents a detailed derivation of the equations of motion
for a multiple elastic pendulum with $N$ point masses. It is shown
that particular cases, such as the double pendulum, can be recovered
from the general equations of motion by applying the appropriate conditions.
The equations of motion for small oscillations are also obtained,
demonstrating in this case the uncoupled behavior between angular
and extension coordinates. Relations between the natural frequencies
of oscillation for the linear case are also established.

\section{Annex A}

The entries comprising each one of the matrices and vectors in equation
\ref{eq:MatrixForm} are detailed next\footnote{The following notation is used: $\underset{i\rightarrow j}{m}=\sum_{r=i}^{j}m_{r}$}:
\begin{equation}
\mathbf{M}_{\theta\theta_{i,j}}=\underset{\max\left(i,j\right)\rightarrow N}{m}\left(l_{i}+u_{i}\right)\left(l_{j}+u_{j}\right)\cos\left(\theta_{i}-\theta_{j}\right)
\end{equation}
\begin{equation}
\mathbf{M}_{\theta u_{i,j}}=\underset{\max\left(i,j\right)\rightarrow N}{m}\left(l_{i}+u_{i}\right)\sin\left(\theta_{j}-\theta_{i}\right)
\end{equation}
\begin{equation}
\mathbf{M}_{uu_{i,j}}=\underset{\max\left(i,j\right)\rightarrow N}{m}\cos\left(\theta_{i}-\theta_{j}\right)
\end{equation}

\begin{equation}
\mathbf{N}_{\theta_{i,j}}=2\underset{\max\left(i,j\right)\rightarrow N}{m}\left(l_{i}+u_{i}\right)\cos\left(\theta_{i}-\theta_{j}\right)
\end{equation}
\begin{equation}
\mathbf{N}_{u_{i,j}}=2\underset{\max\left(i,j\right)\rightarrow N}{m}\sin\left(\theta_{i}-\theta_{j}\right)
\end{equation}

\begin{equation}
\mathbf{R}_{\theta_{i,j}}=\underset{\max\left(i,j\right)\rightarrow N}{m}\left(l_{i}+u_{i}\right)\left(l_{j}+u_{j}\right)\sin\left(\theta_{i}-\theta_{j}\right)
\end{equation}

\begin{equation}
\mathbf{R}_{u_{i,j}}=-\underset{\max\left(i,j\right)\rightarrow N}{m}\left(l_{j}+u_{j}\right)\cos\left(\theta_{i}-\theta_{j}\right)
\end{equation}
\begin{equation}
\mathbf{p}_{\theta_{i}}=g\underset{i\rightarrow N}{m}\left(l_{i}+u_{i}\right)\sin\theta_{i}
\end{equation}

\begin{equation}
\mathbf{p}_{u_{i}}=-g\underset{i\rightarrow N}{m}\left(l_{i}+u_{i}\right)\cos\theta_{i}
\end{equation}
where $\mathbf{M}_{u\theta}=\mathbf{M^{T}}_{\theta u}$, also $\mathbf{K}_{u}$
and $\mathbf{C}_{u}$ are a diagonal matrices whose diagonal elements
are: $k_{i}$ and $c_{i}^{u}$, $i=1\ldots n$ respectively. On the
other hand, $\mathbf{C}_{\theta}$ is a tridiagonal symmetric matrix
whose entries are defined below:
\begin{equation}
C_{\theta_{i,j}}=\left\{ \begin{array}{cl}
c_{i}^{\theta}+c_{i+1}^{\theta} & \mathrm{if}\quad i=j\land1\leq i<N\\
c_{N}^{\theta} & \mathrm{if}\quad i=j=N\\
-c_{i+1}^{\theta} & \mathrm{if}\quad j=i+1\\
-c_{i}^{\theta} & \mathrm{if}\quad i=j+1\\
0 & \mathrm{otherwise}
\end{array}\right.
\end{equation}

\section{Annex B}

Matrices in equation \ref{eq:Linear_eq_Matrix} are defined next:

\begin{equation}
\mathbf{\mathbb{M}}_{\theta}=\left(\begin{array}{ccccc}
\underset{1\rightarrow N}{m}l_{1}^{2} & \underset{2\rightarrow N}{m}l_{1}l_{2} & \underset{3\rightarrow N}{m}l_{1}l_{3} & \cdots & m_{N}l_{1}l_{N}\\
\underset{2\rightarrow N}{m}l_{1}l_{2} & \underset{2\rightarrow N}{m}l_{2}^{2} & \underset{3\rightarrow N}{m}l_{2}l_{3} & \cdots & m_{N}l_{2}l_{N}\\
\underset{3\rightarrow N}{m}l_{1}l_{3} & \underset{3\rightarrow N}{m}l_{2}l_{3} & \underset{3\rightarrow N}{m}l_{3}^{2} & \cdots & m_{N}l_{3}l_{N}\\
\vdots & \vdots & \vdots & \ddots & \vdots\\
m_{N}l_{1}l_{N} & m_{N}l_{2}l_{N} & m_{N}l_{3}l_{N} & \cdots & m_{N}l_{N}^{2}
\end{array}\right)
\end{equation}
\[
\mathbf{\mathbb{K}}_{\theta}=g\left(\begin{array}{ccccc}
\underset{1\rightarrow N}{m}l_{1}\\
 & \underset{2\rightarrow N}{m}l_{2}\\
 &  & \underset{3\rightarrow N}{m}l_{3}\\
 &  &  & \ddots\\
 &  &  &  & m_{N}l_{N}
\end{array}\right)
\]
\begin{equation}
\mathbf{\mathbb{M}}_{u}=\left(\begin{array}{ccccc}
\underset{1\rightarrow N}{m} & \underset{2\rightarrow N}{m} & \underset{3\rightarrow N}{m} & \cdots & m_{N}\\
\underset{2\rightarrow N}{m} & \underset{2\rightarrow N}{m} & \underset{3\rightarrow N}{m} & \cdots & m_{N}\\
\underset{3\rightarrow N}{m} & \underset{3\rightarrow N}{m} & \underset{3\rightarrow N}{m} & \cdots & m_{N}\\
\vdots & \vdots & \vdots & \ddots & \vdots\\
m_{N} & m_{N} & m_{N} & \cdots & m_{N}
\end{array}\right)
\end{equation}
\begin{equation}
\mathbb{\mathbf{\mathbb{K}}}_{u}=\left(\begin{array}{ccccc}
k_{1}\\
 & k_{2}\\
 &  & k_{3}\\
 &  &  & \ddots\\
 &  &  &  & k_{N}
\end{array}\right)
\end{equation}

\bibliographystyle{ieeepes}
\bibliography{MultipleElasticPendulum}

@Book{Meirovitch1967,
  Title                    = {Analytical Methods in Vibrations},
  Author                   = {Leonard Meirovitch},
  Publisher                = {Macmillan Publishing Co},
  Year                     = {1967},
  Quality                  = {1}
}

@Article{Braun2003,
  author       = {Braun, M.},
  date         = {2003},
  journaltitle = {Archive of Applied Mechanics},
  title        = {On some properties of the multiple pendulum},
  doi          = {10.1007/s00419-002-0263-4},
  issn         = {1432-0681},
  number       = {11},
  pages        = {899--910},
  url          = {https://doi.org/10.1007/s00419-002-0263-4},
  volume       = {72},
  refid        = {Braun2003},
}

@Article{Ohlhoff2000,
  author       = {Ohlhoff, A. and Richter, P.H.},
  date         = {2000},
  journaltitle = {ZAMM - Journal of Applied Mathematics and Mechanics / Zeitschrift für Angewandte Mathematik und Mechanik},
  title        = {Forces in the Double Pendulum},
  doi          = {https://doi.org/10.1002/1521-4001(200008)80:8<517::AID-ZAMM517>3.0.CO;2-1},
  eprint       = {https://onlinelibrary.wiley.com/doi/pdf/10.1002/1521-4001%28200008%2980%3A8%3C517%3A%3AAID-ZAMM517%3E3.0.CO%3B2-1},
  number       = {8},
  pages        = {517-534},
  url          = {https://onlinelibrary.wiley.com/doi/abs/10.1002/1521-4001%28200008%2980%3A8%3C517%3A%3AAID-ZAMM517%3E3.0.CO%3B2-1},
  volume       = {80},
}

@Article{Rafat2009,
  author       = {Rafat, M. Z. and Wheatland, M. S. and Bedding, T. R.},
  date         = {2009-03},
  journaltitle = {American Journal of Physics},
  title        = {Dynamics of a double pendulum with distributed mass},
  doi          = {10.1119/1.3052072},
  eprint       = {https://pubs.aip.org/aapt/ajp/article-pdf/77/3/216/13067984/216_1_online.pdf},
  issn         = {0002-9505},
  number       = {3},
  pages        = {216-223},
  url          = {https://doi.org/10.1119/1.3052072},
  volume       = {77},
}

@Article{Stachowiak2006,
  author       = {Tomasz Stachowiak and Toshio Okada},
  date         = {2006},
  journaltitle = {Chaos, Solitons & Fractals},
  title        = {A numerical analysis of chaos in the double pendulum},
  doi          = {https://doi.org/10.1016/j.chaos.2005.08.032},
  issn         = {0960-0779},
  number       = {2},
  pages        = {417-422},
  url          = {https://www.sciencedirect.com/science/article/pii/S0960077905006703},
  volume       = {29},
}

@Article{Awrejcewicz2007,
  author       = {Awrejcewicz, Jan and Kudra, Grzegorz and Wasilewski, Grzegorz},
  date         = {2007},
  journaltitle = {Nonlinear Dynamics},
  title        = {Experimental and numerical investigation of chaotic regions in the triple physical pendulum},
  doi          = {10.1007/s11071-007-9235-0},
  issn         = {1573-269X},
  number       = {4},
  pages        = {755--766},
  url          = {https://doi.org/10.1007/s11071-007-9235-0},
  volume       = {50},
  refid        = {Awrejcewicz2007},
}

@Misc{Yesilyurt2020,
  author      = {Boran Yesilyurt},
  date        = {2020},
  title       = {Equations of Motion Formulation of a Pendulum Containing N-point Masses},
  eprint      = {1910.12610},
  eprintclass = {physics.class-ph},
  eprinttype  = {arXiv},
  url         = {https://arxiv.org/abs/1910.12610},
}

@Article{Szuminski2024,
  author       = {Wojciech Szumi?ski and Andrzej J. Maciejewski},
  date         = {2024},
  journaltitle = {Journal of Sound and Vibration},
  title        = {Dynamics and non-integrability of the double spring pendulum},
  doi          = {https://doi.org/10.1016/j.jsv.2024.118550},
  issn         = {0022-460X},
  pages        = {118550},
  url          = {https://www.sciencedirect.com/science/article/pii/S0022460X24003134},
  volume       = {589},
}

@Article{Belendez2012,
  author       = {A. Belendez and E. Arribas and M. Ortuno and S. Gallego and A. Marquez and I. Pascual},
  date         = {2012},
  journaltitle = {Computers & Mathematics with Applications},
  title        = {Approximate solutions for the nonlinear pendulum equation using a rational harmonic representation},
  doi          = {https://doi.org/10.1016/j.camwa.2012.01.007},
  issn         = {0898-1221},
  number       = {6},
  pages        = {1602-1611},
  url          = {https://www.sciencedirect.com/science/article/pii/S089812211200017X},
  volume       = {64},
}

@Article{Ochs2011,
  author       = {Ochs, Karlheinz},
  date         = {2011-02},
  journaltitle = {European Journal of Physics},
  title        = {A comprehensive analytical solution of the nonlinear pendulum},
  doi          = {10.1088/0143-0807/32/2/019},
  number       = {2},
  pages        = {479},
  url          = {https://doi.org/10.1088/0143-0807/32/2/019},
  volume       = {32},
}

\end{document}